\documentstyle[pra,aps,amssymb]{revtex}

\begin{document}

\tighten
\draft

\title{Multivariable continuous Hahn and Wilson polynomials related to\\
integrable difference systems\thanks{To appear in J. Phys. A.: Math. Gen.}}

\author{J. F. van Diejen\thanks{Supported by the Japan Society for the
Promotion of Science (JSPS).}}

\address{Department of Mathematical Sciences,
University of Tokyo,\\ Hongo 7-3-1, Bunkyo-ku, Tokyo 113, Japan}

\date{May 1995}

\maketitle

\begin{abstract}
Multivariable generalizations of the continuous Hahn and Wilson polynomials
are introduced as eigenfunctions of rational Ruijsenaars type
difference systems with an external field.
\end{abstract}


In a by now famous pioneering paper, Calogero studied (essentially)
the spectrum
and eigenfunctions of a quantum system of $N$ one-dimensional particles
placed
in a harmonic well and interacting by means of an inverse square pair
potential \cite{cal:solution}.
The spectrum of the system, which is discrete due to the harmonic
confinement, turns out to
be remarkably simple: it coincides with that of {\em noninteracting}
particles in a harmonic well up to an overall shift of the energy.
The structure of the corresponding eigenfunctions is also quite simple:
they are the product of a factorized (Jastrow type) ground state wave
function and certain symmetric polynomials. Recently, a rather explicit
construction of these polynomials was given in terms of raising and
lowering operators \cite{bri-han-vas:explicit}.

Some time after its introduction, Olshanetsky and Perelomov realized that
the Calogero system can be naturally generalized within a Lie-theoretic
setting, such that for each (normalized) root system
there exists an associated Calogero type quantum model
\cite{ols-per:quantum}. From this viewpoint, the original model corresponds
to the root system $A_{N-1}$. If one restricts attention to classical
(i.e., non exceptional) normalized root systems, then the Hamiltonians of
the corresponding Calogero models split up in two types:

\noindent {\em Type A: root system $A_{N-1}$}
\begin{equation}
H_{Cal,A} = -\sum_{1\leq j\leq N} \partial_j^2 +\;
g(g-1) \sum_{1\leq j\neq k\leq N} (x_j-x_k)^{-2} \;\; +\,
\omega_0^2 \sum_{1\leq j\leq N} x_j^2 , \label{HcalA}
\end{equation}
\noindent {\em Type B: root systems $B_N$ and $D_N$ ($g_0=0 $)}
\begin{eqnarray}
H_{Cal,B} &=& -\sum_{1\leq j\leq N} \partial_j^2 +\;
g(g-1) \sum_{1\leq j\neq k\leq N}
\left( (x_j-x_k)^{-2} + (x_j+x_k)^{-2} \right) \nonumber \\
& & \;\;\;\;\;\;\;\;\;\;\;\;\;\;\;\;\;\;\;\;\;\;\;\;\;\;\;\;\;\;\;
+ \sum_{1\leq j\leq N}  \left( g_0(g_0-1)\, x_j^{-2}
                      +\; \omega_0^2\, x_j^2  \right)   \label{HcalB}
\end{eqnarray}
($\partial_j = \partial/\partial x_j$).
For the very special case $N=1$, (the polynomial parts of) the
eigenfunctions
of these Hamiltonians amount to Hermite polynomials (type A) and
Laguerre polynomials (type B).
If the external harmonic field is switched off, i.e. for $\omega_0 =0$,
then the spectrum of the Hamiltonian becomes continuous (assuming
$g,g_0 \geq 0$), and the
eigenfunctions are no longer polynomials but give rise to multivariable
families of Bessel type functions \cite{jeu:dunkl}.

Several years ago, Ruijsenaars introduced a relativistic generalization
of the type A model without external field \cite{rui:complete}. The
Hamiltonian of this relativistic system is given by an analytic
difference operator that goes over in $H_{Cal,A}$ (\ref{HcalA})
(with $\omega_0 =0$) after sending the step size to zero.
(This transition may be interpreted as the nonrelativistic limit.)
More recently we found similar difference versions for the type A model
with $\omega_0 \neq 0$ and for the type B model \cite{die:difference}.
In this letter we study the eigenfunctions of these two difference
models. For the type A difference version this will lead to a
multivariable generalization of the continuous Hahn
polynomials \cite{ata-sus:hahn,ask:continuous}, whereas
for type B we will find a multivariable version of the Wilson
polynomials \cite{wil:some,ask-wil:set}.
By sending the step size to zero, we recover the spectrum and
eigenfunctions of
the Calogero Hamiltonians $H_{Cal,A}$ (\ref{HcalA}) and
$H_{Cal,B}$ (\ref{HcalB}).

\pagebreak
{\bf Type A: continuous Hahn case.}
The Hamiltonian of the (rational) difference Ruijsenaars system with
external field
is given by the second order difference operator \cite{die:difference}
\begin{eqnarray}
H_A &=& \sum_{1\leq j\leq N} \Bigl(
w_+^{1/2}(x_j)\prod_{k\neq j} v^{1/2} (x_j-x_k)\: e^{-i\partial_j}\:
\prod_{k\neq j} v^{1/2} (x_k-x_j) w_-^{1/2}(x_j)  \nonumber \\
& &\;\;\;\;\;\;\;\;
+ w_-^{1/2}(x_j)\prod_{k\neq j} v^{1/2} (x_k-x_j)\: e^{i\partial_j}\:
\prod_{k\neq j} v^{1/2} (x_j-x_k) w_+^{1/2}(x_j)  \nonumber \\
& &\;\;\;\;\;\;\;\; -w_+(x_j)\prod_{k\neq j} v (x_j-x_k)
-w_-(x_j)\prod_{k\neq j} v (x_k-x_j) \Bigr)  \label{HA}
\end{eqnarray}
where $(e^{\pm i\partial_j}\Psi )(x_1,\ldots ,,x_N)=
\Psi (x_1,\ldots ,x_{j-1},x_j\pm i ,x_{j+1},\ldots ,x_N)$ and
\begin{equation}\label{vwA}
v(z)=1+g/(iz) , \;\;\;
w_+(z)=(a_+ +iz)(b_+ +iz), \;\;\; w_-(z)=(a_- -iz)(b_- -iz).
\end{equation}
We will assume
\begin{equation}\label{hermitian}
g\geq 0, \;\;\;\;\; a_-=\overline{a}_+, \;\;\;\;\; b_-=\overline{b}_+,
\;\;\;\;\; \text{Re}(a_+,a_-,b_+,b_-)>0,
\end{equation}
which ensures, in particular, that the Hamiltonian is formally self-adjoint.
In order to solve the eigenvalue problem for
$H_A$ (\ref{HA})-(\ref{hermitian}) in the Hilbert space of square integrable
permutation invariant functions,
we introduce the weight function
\begin{equation}\label{weightA}
\Delta_A= \prod_{1\leq j\neq k\leq N}\!\!\!
\frac{\Gamma (g+i(x_j-x_k))}{\Gamma (i(x_j-x_k))}\;\;
\prod_{1\leq j\leq N}\!\! \Gamma (a_+ +ix_j)\Gamma (b_+ +ix_j)
\Gamma (a_- -ix_j) \Gamma (b_- -ix_j) .
\end{equation}
Condition~(\ref{hermitian}) implies that $\Delta_A$ is positive.
Since the transformed operator
\begin{eqnarray}
{\cal H}_A &=& \Delta_A^{-1/2}H_A\Delta_A^{1/2} \nonumber \\
&=& \sum_{1\leq j\leq N} \Bigl(
w_+(x_j)\prod_{k\neq j} v (x_j-x_k) \left( e^{-i\partial_j}-1 \right) +
\nonumber \\
& & \;\;\;\;\;\;\;\;\;\;
w_-(x_j)\prod_{k\neq j} v (x_k-x_j) \left( e^{i\partial_j}-1 \right) \Bigr)
\label{HtA}
\end{eqnarray}
clearly annihilates constant functions, it follows that $\Delta_A^{1/2}$
is an eigenfunction of $H_A$ with eigenvalue
zero. This eigenfunction corresponds to the ground state. The excited
states are a product of $\Delta_A^{1/2}$ and symmetric polynomials
associated with the weight function $\Delta_A$ (\ref{weightA}).

Specifically, one has
\begin{equation}
H_A \,\Psi_{\vec{n}} =E_A(\vec{n})\, \Psi_{\vec{n}},\;\;\;\;\;\;
\vec{n}\in {\Bbb Z}^N, \;\; n_1\geq n_2\geq \cdots \geq n_N \geq 0,
\end{equation}
with eigenvalues reading
\begin{equation}\label{eigwA}
E_A(\vec{n}) =
\sum_{1\leq j\leq N} n_j (\, n_j+a_+ +a_-  +b_+ +b_- -1+ 2(N-j)g  \, ),
\end{equation}
and eigenfunctions of the form
\begin{equation}
\Psi_{\vec{n}}(\vec{x}) =\Delta_A^{1/2}\, p_{\vec{n}, A}(\vec{x}),
\end{equation}
where $p_{\vec{n}, A}(\vec{x})$ denotes the symmetric polynomial determined
by the conditions
\begin{itemize}
\item[A.1]
$\displaystyle p_{\vec{n}, A}(\vec{x}) = m_{\vec{n}}(\vec{x})
  + \sum_{\vec{n}^\prime < \vec{n}} c_{\vec{n},\vec{n}^\prime}\:
               m_{\vec{n}^\prime}(\vec{x}),\;\;\;\;\;\;\;\;\;\;
c_{\vec{n},\vec{n}^\prime}\in {\Bbb C}$;

\item[A.2]
$\displaystyle
\int_{{\Bbb R}^N} p_{\vec{n}, A}(\vec{x})\:
\overline{m_{\vec{n}^\prime}(\vec{x})}\: \Delta_A \;dx_1\cdots dx_N
=0\;\;\;\;\;\;
\text{if}\;\;\;\; \vec{n}^\prime < \vec{n}$.
\end{itemize}
Here the functions $m_{\vec{n}}(\vec{x})$ denote the basis of monomial
symmetric functions
\begin{equation}
m_{\vec{n}}(\vec{x}) = \sum_{\vec{n}^\prime \in S_N (\vec{n})}
x_1^{n_1^\prime}\cdots x_N^{n_N^\prime},\;\;\;\;\;\;
\vec{n}\in {\Bbb Z}^N, \;\; n_1\geq n_2\geq \cdots \geq n_N \geq 0,
\end{equation}
and the partial order of the basis elements is defined by
\begin{equation}\label{order}
\vec{n}^\prime \leq \vec{n} \;\;\;\;\;\; \text{iff}\;\;\;\;\;\;
\sum_{1\leq j\leq k} n_j^\prime \leq \sum_{1\leq j\leq k} n_j\;\;\;
\text{for}\;\; k=1,\ldots ,N
\end{equation}
($\vec{n}^\prime < \vec{n}$ if $\vec{n}^\prime \leq \vec{n}$ and
$\vec{n}^\prime \neq \vec{n}$).

The proof of the above statement boils down to showing that the polynomials
$p_{\vec{n}, A}(\vec{x})$ are eigenfunctions of the transformed operator
${\cal H}_A$ (\ref{HtA}).
This follows from the fact that ${\cal H}_A$
is both triangular with respect to the monomial basis:
\begin{equation}\label{triangular}
({\cal H}_A \, m_{\vec{n}}) (\vec{x}) =
\sum_{\vec{n}^\prime\leq \vec{n}} [{\cal H}_A]_{\vec{n},\vec{n}^\prime}\:
m_{\vec{n}^\prime}(\vec{x}) ,
\end{equation}
and symmetric with respect to the $L^2$ inner product
with weight function $\Delta_A$ (\ref{weightA}).
The eigenvalues $E_A(\vec{n})$ (\ref{eigwA})
are obtained by computing the diagonal matrix elements
$[{\cal H}_A]_{\vec{n},\vec{n}}$, i.e., the
leading coefficients in Expansion~(\ref{triangular}) of
$({\cal H}_A \, m_{\vec{n}}) (\vec{x})$ in monomial symmetric functions
$m_{\vec{n}^\prime}(\vec{x})$.

The polynomial $p_{\vec{n}, A}(\vec{x})$ by definition amounts to
$m_{\vec{n}}(\vec{x})$ minus
its orthogonal projection in $L^2({\Bbb R}^N, \Delta_A dx_1\cdots dx_N)$
onto $\text{span}\{ m_{\vec{n}^\prime} \}_{\vec{n}^\prime < \vec{n}}$.
For $N=1$ these polynomials reduce to (monic) continuous Hahn
polynomials \cite{ata-sus:hahn,ask:continuous}.

{\bf Type B: Wilson case.}
Multivariable Wilson polynomials are obtained much in the same manner as
their continuous Hahn counterparts, except that in addition to
being permutation
symmetric now everything also becomes even in $x_j$, $j=1,\ldots ,N$.
The Hamiltonian of the type B version of
the difference Ruijsenaars system reads \cite{die:difference}
\begin{equation}\label{HB}
H_B = \sum_{1\leq j\leq n} \left(
V^{1/2}_{+j}\, e^{-i\partial_j}\, V^{1/2}_{-j}\: +\:
V^{1/2}_{-j}\, e^{i\partial_j}\, V^{1/2}_{+j}\:
-V_{+j}\, -V_{-j} \right) ,
\end{equation}
with
\begin{eqnarray}
V_{\pm j}\! &=&\! w(\pm x_j)
              \prod_{k\neq j} v(\pm x_j+x_k)\, v(\pm x_j-x_k) , \\
v(z)\! &=&\!  1+g/(iz), \;\;\;\;\;\;\;
w(z)=\frac{(a+iz)(b+iz)(c+iz)(d+iz)}{2iz (2iz+1)},
\end{eqnarray}
and
\begin{equation}
 g\geq 0,\;\;\;\;\;\;\;\;\;\; a,b,c,d >0.
\end{equation}
One now has
\begin{equation}
H_B\,\Psi_{\vec{n}} =E_B(\vec{n})\, \Psi_{\vec{n}},\;\;\;\;\;\;
\vec{n}\in {\Bbb Z}^N, \;\; n_1\geq n_2\geq \cdots \geq n_N \geq 0,
\end{equation}
with
\begin{equation}
E_B(\vec{n}) = \sum_{1\leq j\leq N} n_j (\, n_j+a+b+c+d-1+ 2(N-j)g  \, ),
\;\;\;\;\;\;\;
\Psi_{\vec{n}}(\vec{x}) =\Delta_B^{1/2}\, p_{\vec{n},B}(\vec{x}^2),
\end{equation}
where
\begin{eqnarray}
\Delta_B &=&
\prod_{1\leq j\neq k\leq N}
\left| \frac{\Gamma (g+i(x_j+x_k))\, \Gamma (g+ i(x_j-x_k))}
     {\Gamma (i(x_j+x_k))\, \Gamma (i(x_j-x_k)) } \right| \nonumber \\
& & \times \prod_{1\leq j\leq N} \left|
\frac{\Gamma (a+ix_j)\Gamma (b+ix_j) \Gamma (c+ix_j) \Gamma (d+ix_j)}
     {\Gamma (2ix_j)} \right|^2 , \label{weightB}
\end{eqnarray}
and $p_{\vec{n},B}(\vec{x}^2)$ is the even symmetric polynomial
determined by the conditions
\begin{itemize}
\item[B.1]
$\displaystyle p_{\vec{n},B}(\vec{x}^2) = m_{\vec{n}}(\vec{x}^2)
  + \sum_{\vec{n}^\prime < \vec{n}} c_{\vec{n},\vec{n}^\prime}\:
               m_{\vec{n}^\prime}(\vec{x}^2),\;\;\;\;\;\;\;\;\;\;
c_{\vec{n},\vec{n}^\prime}\in {\Bbb C}$;

\item[B.2]
$\displaystyle
\int_{{\Bbb R}^N} p_{\vec{n},B}(\vec{x}^2)\:
\overline{m_{\vec{n}^\prime}(\vec{x}^2)}\: \Delta_B \;dx_1\cdots dx_N
=0\;\;\;\;\;\;
\text{if}\;\;\;\; \vec{n}^\prime < \vec{n}$.
\end{itemize}
Here the functions $m_{\vec{n}}(\vec{x}^2)$ stand for the basis of even
symmetric monomials
\begin{equation}
m_{\vec{n}}(\vec{x}^2) = \sum_{\vec{n}^\prime \in S_N (\vec{n})}
x_1^{2n_1^\prime}\cdots x_N^{2n_N^\prime},\;\;\;\;\;\;
\vec{n}\in {\Bbb Z}^N, \;\; n_1\geq n_2\geq \cdots \geq n_N \geq 0,
\end{equation}
and the partial order of the basis elements is the same as before
(see (\ref{order})).

The polynomials $p_{\vec{n},B}(\vec{x}^2)$ are of course again
eigenfunctions of the transformed operator
\begin{eqnarray}
{\cal H}_B &=& \Delta_B^{-1/2}H_B\Delta_B^{1/2} \nonumber \\
&=& \sum_{1\leq j\leq N} \Bigl(
w(x_j)\prod_{k\neq j} v (x_j+x_k)\, v (x_j-x_k)
\left( e^{-i\partial_j}-1 \right) +
\nonumber \\
& & \;\;\;\;\;\;\;\;\;\;
w(-x_j)\prod_{k\neq j} v(-x_j+x_k)\, v (-x_j-x_k) \left( e^{i\partial_j}-1
\right) \Bigr) . \label{HtB}
\end{eqnarray}
For $N=1$ they reduce to (monic) Wilson polynomials
\cite{wil:some,ask-wil:set}.

{\bf Transition to the Calogero system.}
If we substitute $x_j\rightarrow \beta^{-1} x_j$ (so
$\partial_j\rightarrow \beta\partial_j$) and
$a_+,a_-\rightarrow (\beta^2\omega_0)^{-1}$, \ \
$b_+,b_-\rightarrow (\beta^2\omega_0^\prime )^{-1}$ in our type A difference
Hamiltonian and multiply by $\beta^2\omega_0\omega_0^\prime$,
then we arrive at a Hamiltonian of the form $H_A$ (\ref{HA}) with
$\exp (\pm i\partial_j)$ replaced by $\exp (\pm i\beta \partial_j)$ and
\begin{equation}
v(z) = 1 +\beta g/(iz),\;\;\;\;\;\;\;\;\;\;\;\;\;
w_\pm(z) = \beta^{-2} (1\pm i\beta\omega_0\, z)
                      (1\pm i\beta\omega_0^\prime \, z) .
\end{equation}
By sending the step size $\beta$ to zero the difference operator
goes over in a differential operator of the form
$H_{Cal,A}-\varepsilon_{0,A}$,
where $H_{Cal,A}$ is given by
(\ref{HcalA}) with $\omega_0$ replaced by $\omega_0+\omega_0^\prime$, and
$\varepsilon_{0,A}$ is a constant with value
\begin{equation}
\varepsilon_{0,A}= (\omega_0+\omega_0^\prime) N(1+(N-1)g).
\end{equation}
In this limit the weight function $\Delta_A$ (\ref{weightA}) goes (after
dividing by a divergent numerical factor) over in
\begin{equation}\label{weightCA}
\Delta_{Cal,A} =\prod_{1\leq j\neq k\leq N} |x_j-x_k|^g\,
              \prod_{1\leq j\leq N} e^{-(\omega_0+\omega_0^\prime)\, x^2}.
\end{equation}
Thus, we recover the spectrum and eigenfunctions of the type A Calogero
system:
\begin{equation}
H_{Cal,A} \,\Psi_{\vec{n}} =E_{Cal,A}(\vec{n})\, \Psi_{\vec{n}},
\;\;\;\;\;\;
\vec{n}\in {\Bbb Z}^N, \;\; n_1\geq n_2\geq \cdots \geq n_N \geq 0,
\end{equation}
with
\begin{equation}
E_{Cal, A}(\vec{n}) = \varepsilon_{0,A}+
2(\omega_0+\omega_0^\prime)\sum_{1\leq j\leq N} n_j,
\;\;\;\;\;\;\;\;\;
\Psi_{\vec{n}}(\vec{x}) =\Delta_{Cal,A}^{1/2}\: p_{\vec{n},Cal, A}(\vec{x}),
\end{equation}
where $p_{\vec{n}, Cal, A}(\vec{x})$ denotes the symmetric polynomial
determined by Conditions A.1, A.2 with $\Delta_{A}$ (\ref{weightA})
replaced by
$\Delta_{Cal,A}$ (\ref{weightCA}).

For type B, the transition to the Calogero system is quite similar. The
substitution $x_j\rightarrow \beta^{-1} x_j$
($\partial_j\rightarrow \beta\partial_j$), $a\rightarrow g_0$,
$b\rightarrow g_0^\prime +1/2$, $c\rightarrow (\beta^2\omega_0)^{-1}$,
$d\rightarrow (\beta^2\omega_0^\prime)^{-1}$, and
$H_{B}\rightarrow 4\beta^2 \omega_0 \omega_0^\prime H_B$ leads to a
Hamiltonian of the form $H_B$ (\ref{HB}) with
$\exp (\pm i\partial_j)$ replaced by $\exp (\pm i\beta \partial_j)$ and
\begin{eqnarray}
v(z) &=& 1+\beta g/(iz),\;\;\;\;\;\;\;\; \nonumber \\
w(z) &=& \beta^{-2} (1+\frac{\beta g_0}{iz})
                  (1+\frac{\beta g_0^\prime}{(iz+\beta/2)})
                  (1+i\beta\omega_0\, z)
                  (1+ i\beta\omega_0^\prime \, z) .
\end{eqnarray}
For $\beta\rightarrow 0$ one now obtains a differential operator of the form
$H_{Cal,B}-\varepsilon_{0,B}$, where
$H_{Cal,B}$ is given by (\ref{HcalB}) with $g_0$ replaced by
$g_0+g_0^\prime$
and $\omega_0$ by $\omega_0+\omega_0^\prime$, and the constant
$\varepsilon_{0,B}$ reads
\begin{equation}
\varepsilon_{0,B} =
 (\omega_0+\omega_0^\prime) N (1+2(N-1)g+2(g_0+g_0^\prime)).
\end{equation}
In the limit (and after dividing by a divergent numerical factor)
the type B weight function goes over in
\begin{equation}\label{weightCB}
\Delta_{Cal,B} = \prod_{1\leq j\neq k\leq N} |x_j+x_k|^g |x_j-x_k|^g\,
               \prod_{1\leq j\leq N}
      |x_j|^{2(g_0+g_0^\prime)}\, e^{-(\omega_0+\omega_0^\prime)x_j^2},
\end{equation}
and the eigenfunctions become:
\begin{equation}
H_{Cal,B} \,\Psi_{\vec{n}} =
E_{Cal, B}(\vec{n})\, \Psi_{\vec{n}},\;\;\;\;\;\;
\vec{n}\in {\Bbb Z}^N, \;\; n_1\geq n_2\geq \cdots \geq n_N \geq 0,
\end{equation}
with
\begin{equation}
E_{Cal, B}(\vec{n}) = \varepsilon_{0,B}+
 4(\omega_0+\omega_0^\prime)\sum_{1\leq j\leq N} n_j,
\;\;\;\;\;\;\;\;\;
\Psi_{\vec{n}}(\vec{x}) =
\Delta_{Cal,B}^{1/2}\: p_{\vec{n},Cal, B}(\vec{x}^2),
\end{equation}
where $p_{\vec{n}, Cal, B}(\vec{x}^2)$ denotes the even symmetric polynomial
determined by the Conditions B.1, B.2 with $\Delta_{B}$ (\ref{weightB})
replaced by $\Delta_{Cal,B}$ (\ref{weightCB}).

Let us conclude by remarking that both our multivariable
continuous Hahn and Wilson polynomials are limiting cases of a
multivariable version of the Askey-Wilson polynomials \cite{ask-wil:basic}
introduced by Koornwinder \cite{koo:askey} as a generalization
of Macdonald's polynomials associated with the root system $BC_N$
\cite{mac:orthogonal1,mac:orthogonal2}.
Koornwinder's multivariable Askey-Wilson
polynomials are joint eigenfunctions
of an algebra of commuting difference operators with trigonometric
coefficients \cite{die:commuting}. This algebra constitutes a
complete set of quantum integrals for a difference version
of the trigonometric $BC_N$-type Calogero-Sutherland system.
Similar algebras consisting of commuting difference operators that are
simultaneously diagonalized by our multivariable continuous Hahn and Wilson
polynomials can be obtained as rational degenerations.
The difference Hamiltonians considered in this letter are the simplest
(i.e., lowest order) nontrivial operators in these algebras.


\begin{thebibliography}{10}

\bibitem{cal:solution} Calogero F
1971 {\em J. Math. Phys.} {\bf 12} 419-36

\bibitem{bri-han-vas:explicit} Brink L, Hansson T H, and Vasiliev M A
1992 {\em Phys. Lett. B} {\bf 286} 109-11

\bibitem{ols-per:quantum} Olshanetsky M A and Perelomov A M
1983 {\em Phys. Reps.} {\bf 94} 313-404

\bibitem{jeu:dunkl} de Jeu M F E
1993 {\em Invent. Math.} {\bf 113} 147-62

\bibitem{rui:complete} Ruijsenaars S N M
1987 {\em Commun. Math. Phys.} {\bf 110} 191-213

\bibitem{die:difference} van Diejen J F
1995 {\em J. Math. Phys.} {\bf 36} 1299-1323

\bibitem{ata-sus:hahn} Atakishiyev N M and Suslov S K
1985 {\em J. Phys. A: Math. Gen.} {\bf 18} 1583-96

\bibitem{ask:continuous} Askey R
1985 {\em J. Phys. A: Math. Gen.} {\bf 18} L1017-9

\bibitem{wil:some} Wilson J A
1980 {\em SIAM J. Math. Anal.} {\bf 11} 690-701

\bibitem{ask-wil:set} Askey R and Wilson J A
1982 {\em SIAM J. Math. Anal.} {\bf 13} 651-5

\bibitem{ask-wil:basic} Askey R and Wilson J A
1985 {\em Mem. Amer. Math. Soc.} {\bf 54} no. 319

\bibitem{koo:askey} Koornwinder T H
1992 {\em Hypergeometric functions on domains of
positivity, Jack polynomials,
and applications} ed D St P Richards  {\em Contemp. Math.} vol. 138
(Providence, R. I.: Amer. Math. Soc.) pp. 189-204

\bibitem{mac:orthogonal1} Macdonald I G
1990 {\em Orthogonal polynomials:
theory and practice} ed P Nevai {\em NATO ASI Series C 294}
(Dordrecht: Kluwer) pp. 311-8.


\bibitem{mac:orthogonal2}  Macdonald I G 1988 {\em Orthogonal polynomials
associated with root systems} preprint Univ. of London


\bibitem{die:commuting} van Diejen J F
1995 {\em Compos. Math.} {\bf 95}
183-233

\end{thebibliography}
\end{document}